\documentclass[sigconf]{acmart}
\AtBeginDocument{%
  }

\usepackage{subcaption}
\usepackage{amsmath}
\usepackage{booktabs}
\usepackage{multirow}
\usepackage{tikz}
\usepackage{adjustbox}
\usetikzlibrary{shapes,arrows,positioning,fit,backgrounds, calc}

\newcommand{\best}[1]{\textbf{#1}}   
\newcommand{\comparable}[1]{\textit{#1}$^{*}$}   
\setcopyright{cc}
\setcctype{by}
\copyrightyear{2026}
\acmYear{2026}
\acmDOI{10.1145/3799682.3840092}
\acmConference[CIKM '26]{Proceedings of the 35th ACM International Conference on Information and Knowledge Management}{November 07--11, 2026}{Rome, Italy}
\acmBooktitle{Proceedings of the 35th ACM International Conference on Information and Knowledge Management (CIKM '26), November 07--11, 2026, Rome, Italy}
\acmISBN{979-8-4007-2539-5/2026/11}

\begin{document}

\title{Bridging Queries and Brands: Entity Linking for Large-Scale Product Search}

\author{Dong Liu}
\authornote{Both authors contributed equally to this research.}
\email{liuadong@amazon.com}
\affiliation{%
  \institution{Amazon}
  \city{Luxembourg}
  \country{Luxembourg}
}
\author{Sreyashi Nag}
\authornote{Work done while at Amazon.}
\email{sreyanag@amazon.com}
\authornotemark[1]
\affiliation{%
  \institution{Amazon}
  \city{Palo Alto}
  \state{California}
  \country{United States}
}

\renewcommand{\shortauthors}{Dong Liu and Sreyashi Nag}

\begin{abstract}
  Associating user search queries with the correct brand entity is critical for e-commerce product retrieval, yet remains challenging due to the brevity of queries (three to four words on average), their lack of grammatical structure, and a catalog of hundreds of thousands of distinct brands. We formulate this as a brand entity linking task and develop two complementary solutions deployed at scale: (1)~a cascaded pipeline that first detects brand mentions via sequence labeling and then disambiguates against a brand knowledge base, and (2)~a single-stage approach that frames linking as extreme multi-class classification, directly mapping queries to brand identifiers. Through extensive multilingual evaluation (11 languages) and a controlled online experiment, we demonstrate that the proposed methods substantially improve brand recall while maintaining high precision, leading to measurable gains in customer engagement.
\end{abstract}

\ccsdesc[500]{Information systems~Query intent}
\ccsdesc[500]{Computing methodologies~Natural language processing}

\keywords{entity linking, brand recognition, e-commerce search, named entity recognition, extreme classification}



\maketitle
\section{Introduction}

Manufacturer identity is one of the most important attributes identified in search queries alongside product-type in e-commerce~\citep{nag2023qau, Luoetal2024}. Correctly recognizing brand names, whether mentioned directly (e.g., through explicit company names) or indirectly (e.g., through product-specific terminology), is an important component of search query understanding and is crucial to providing a good shopping experience to customers.
The current string-based approach to brand identification in search queries presents several challenges: (i) unifying brand name variants across different languages and regions (e.g., a Western brand name written in its original form versus its representation in Asian scripts), (ii) different surface forms for the same brand (e.g., abbreviations versus full names) and (iii) identifying brand relationships between parent and sub-brands (e.g., a parent company and its product line brands).
Therefore, in addition to recognizing the brand names mentioned in the query, it is also important to link them to the corresponding global brand entity.
It would be valuable to unify the concept of brand across different e-commerce stores in a single namespace, i.e., brand entity (identity to each brand itself).
Each brand entity is unique across languages, stores and surface forms. As part of this effort, we aim to recognize the brand entity for branded search queries. We define branded search queries as those queries with clearly specified brand intents, e.g., \textit{<brand> running shoes}. Such queries have a single brand entity as its shopping intent. We view this as an entity linking problem where the brand needs to be identified and subsequently linked to a unique brand entity in our brand database.

Entity linking is the task of recognizing meaningful mentions in a textual context and linking them to a knowledge base~\citep{reinanda2020kg,sevgili2020neural}. In literature, entity linking sometimes assumes that the entity mentions are already given or detected, for example, using a named entity recognition (NER) system \cite{Sukumaretal2024, Tanetal2017, cao-etal-2017-bridge}. However, in this paper, we define entity linking as the combined task of entity recognition and entity disambiguation that operates on the raw textual context of product search queries in e-commerce.

In the e-commerce domain, entity linking is especially important to understand search queries and the customer’s shopping intent. It is used to extract the important attributes in a search query, e.g., brand, product type etc. and link them to known entities in a knowledge base. The most common approach to entity extraction is a two-step process consisting of: i) entity recognition, and ii) entity disambiguation. The goal of entity recognition is to identify meaningful spans from the textual context.  When recognizing multiple entity types, this step also includes tagging the mentioned entity span with its entity type. The recognized spans are usually ambiguous in nature and the entity disambiguation step is then performed to link the span to a knowledge base. Often there are multiple candidate entities in the knowledge base that match the recognized entity. The entity linker generates and ranks these candidates. 

The entity linking of e-commerce queries is challenging since the query length is usually short (approximately three to four words on average~\citep{chen2022queryattr}) and there is limited additional context. Unlike web search, such queries often lack the natural language structure and open source NLP models are unable to handle such query distributions well. Further, 
brand identity is a central attribute in e-commerce search queries and there are hundreds of thousands of unique brands on such services with new ones constantly added. In this work, we proposed different solutions to the brand entity linking problem. Our contributions can be summarized as follows:
\begin{enumerate}
\item We first build a two-stage entity linking model for e-commerce query brand entity prediction, consisting of an NER brand recognition model (we employ the pretrained NER model as shall be explained in \ref{sub-sec-ner}) and a surface form matching step.
\item We explore both lexical and semantic matching techniques for surface form matching and develop a product type-based filtering step for brand entity prediction refinement in massive brand output space.
\item We propose a novel end-to-end extreme multi-class classification model that directly takes the search query as input and predicts the relevant brand entity. The end-to-end model is then integrated with the two-stage model as a fused solution for brand entity linking.
\item We validate the system through extensive offline experiments across 11 languages and a large-scale online A/B test, demonstrating measurable improvements in brand recall and downstream business metrics.
\end{enumerate}

\section{Related Work}
The mainstream approach to entity linking follows a two-step entity design, where entity recognition is separated from entity disambiguation \cite{Tanetal2017, le2019distant, cao-etal-2017-bridge}.
In two-step methods, when a mention span is identified via a Named Entity Recognition (NER) model, there are commonly three ways to generate entity candidates: i) surface form matching; ii) expansion using aliases; iii) prior probability matching. In surface form matching such as \citep{le2019distant, zwicklbauer2016robust, morene2017combine}, a candidate list is generated such that each entity match at least one surface form of the mention. Matching may be done via either exact or fuzzy matching. For expansion with aliases \citep{zhang2019joint}, a dictionary of additional aliases is constructed to group known surface forms and brand variants together. In prior probability methods \citep{spitkovsky12a}, a pre-calculated prior probability of corresponding entities for given certain mentions is used to infer the possible entity candidates for a given query. After the candidate generation in the two-step methods, the entity candidates are usually ranked according to their relevance to the mention input, where the relevance can also take into account the context of the mention. Candidate ranking can be done via computing the Levenshtein distance or via vector product in a semantically enriched embedding space where both entity and mention are represented by semantic vectors \citep{logeswaran-etal-2019-zero,huang2015leverageing, cao-etal-2017-bridge}.


Recent entity linking approaches have also performed joint entity recognition and disambiguation. This includes approaches that perform both steps simultaneously \cite{peters2019knowledge, sorokin2018mixing}, allowing for interaction and shared knowledge between the two steps. Other approaches eliminate the two-step process entirely and tackle this problem as an end-to-end task, directly predicting the entity from the input textual context. For example, \citet{kolitsas2018end} formulate the task as a sequence labeling problem, where each token in the text is assigned an entity link or a NIL class (a none class indicating no entity). Thus, they perform token classification over the entire entity vocabulary. Yet another approach retrieves entities by generating their unique names, left to right, token-by-token in an autoregressive fashion and conditioned on the context~\citep{cao2021autoregressive}. Furthermore, \citet{li2020efficient} present a more advanced bi-encoder entity linking algorithm by using two encoders to embed the input text and the entity descriptions, where the entity candidates are obtained by relevance score based on comparing the embeddings. 

Entity linking in e-commerce contexts presents distinct challenges that differentiate it from traditional (e.g. Wikipedia) approaches. E-commerce queries are typically short and lack the lexical context that traditional entity linking systems rely upon, making the task more restrictive than standard knowledge base linking scenarios \cite{Manchandaetal2020}. The fundamental task could involve extracting query attribute values, which requires both detecting named entities in search queries as diverse surface form attribute values and normalizing them into canonical forms to handle misspelling and abbreviation problems \cite{Zhangetal2021}. As pointed out in \cite{Ricatteetal2023}, queries are often short, noisy, and ambiguous in e-commerce. 
Recent end-to-end architectures such as REXEL~\citep{kroll2024rexel} and LLM-driven approaches~\citep{wang2025aelc} demonstrate progress on unified entity linking, but remain evaluated primarily on news and Wikipedia domains rather than the short-query, large-catalog setting of product search.
Early approaches to e-commerce entity recognition focused on building CRF taggers for recognizing basic entity types such as brands, types, and models in shopping queries, with studies showing promising results when large knowledge bases are available \cite{Kozarevaetal2016,Manshadietal2009}. Parallel efforts have included using Hearst patterns to learn semantic lexicons, though these methods cannot recognize all product entities in queries \cite{Manchandaetal2020}.
The practical applications of e-commerce entity linking are extensive, supporting product retrieval and ranking systems, query rewriting for better user intent understanding, and improving search results quality \cite{Zhangetal2021, Wenetal2019, GuisadoGamezetal2021}. Given the critical role of entity linking in e-commerce, this work focuses on building a systematic solution for brand entity linking in product item shopping. 

\section{Preliminaries}
We introduce three preliminary models that are applied as part of our proposed methods. 

\subsection{MetaTS-NER}\label{sub-sec-ner}
One of our baseline solutions relies on named-entity recognition (NER). Therefore we introduce the NER model used in this work. We utilize the MetaTS-NER \citep{Li2021Meta} model for brand string detection. MetaTS-NER is a multilingual DistilBERT based model for sequential labeling.

The model formulates the problem as a token-level multi-class classification task. For a given input query, the text is first tokenized, e.g., "red nike shoes" becomes \texttt{["red", "ni", "\#\#ke", "sh", "\#\#oes"]}. Special tokens \texttt{[CLS]} and \texttt{[SEP]} are added as start and end markers. These tokens are then passed to the DistilBERT model to generate contextual embeddings for each token. The token embeddings are fed to a multi-class classifier to predict the label of each token. 


The model is trained on internal data to predict attribute type annotations—such as product type, brand, size, etc.—for product search queries. Our initial two-stage brand entity linking framework depends on this model, as it uses the brand name annotations from the queries.


\subsection{Q2PT}
Our solutions rely on product type intents of search queries to disambiguate brand entity linking. We thus need a product type classification model to parse a search query. The classification model that we leverage is an internal model, i.e., query-to-product-type model (Q2PT). The deep learning classification model is a BERT-based search query classification model fitted with an internal dataset. We use the acronym PT for product type. 


\subsection{PECOS}

PECOS (Prediction in Enormous and Correlated Output Spaces)~\citep{yu2022pecos} is a general purpose framework aimed at solving extreme multi-class classification problems. It is particularly well-suited for scenarios with large output spaces (millions to billions of labels) and where many labels have limited training data, which is common in e-commerce applications.
    
    

The mathematical formulation of PECOS can be expressed as follows. Given a training dataset $\{(x_i, y_i) : i = 1, \ldots, n\}$, where $x_i \in \mathbb{R}^d$ is a $d$-dimensional feature vector and $y_i \in \{0,1\}^L$ denotes the relevant labels from an output space $\mathcal{Y} = \{1, \ldots, L\}$, PECOS learns a scoring function $f(x, \ell) : \mathbb{R}^d \times \mathcal{Y} \rightarrow \mathbb{R}$ that maps an input $x$ and a label $\ell$ to a relevance score.

PECOS not only achieves competitive performance in semantically relevant tasks but also offers low inference latency thanks to its hierarchical clustering tree and doubly-sparse data structure~\citep{etter2021accelerating}. For example, the inference latency is less than 5 ms per query for generating predictions from a class space of 100 million products~\citep{chang2021extreme}. In this work, we apply the framework provided by PECOS to train new classification models for our entity disambiguation task.

\section{Brand Entity Linking via Two-Stage Framework}\label{sec_two_stage_framework}
%
%
%

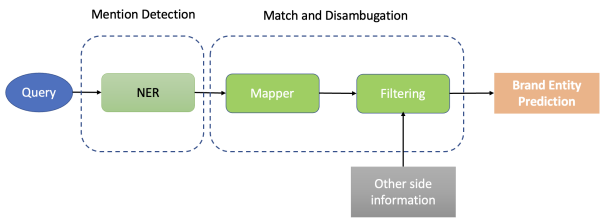
Following the typical two-stage approach in entity linking, i.e., a mention detection step followed by entity disambiguation, we utilize the MetaTS-NER model to build a two-stage framework for brand entity prediction. In the first stage, we extract the brand name from branded search queries using the NER model. Given a query $q$, the first stage applies the NER model $f_{\text{NER}}$ to identify brand mentions:
\begin{equation}
m = f_{\text{NER}}(q)
\end{equation}
where $m$ represents the extracted brand mention.
In the second stage, a matcher function $g$ maps the brand mention to a set of candidate brand entities:
\begin{equation}
E_{\text{ID}} = g(m)
\end{equation}
Finally, a filtering function $h$ that incorporates contextual information (such as product type) selects the most appropriate brand entity:
\begin{equation}
\label{eq:filter-mapping}
e = h(E_{\text{ID}}, q, \text{PT}_q)
\end{equation}
where $\text{PT}_q$ is the product type predicted for query $q$. $e$ is the final predicted brand entity. 
Equation~\ref{eq:filter-mapping} is to address the challenge that a brand name may be mapped to more than one brand entity. 
To address the ambiguity of one-to-many mappings, we implement function $h$ as a filtering step to disambiguate between the multiple brand entity candidates based on the query context. We apply a product-type (PT) based filtering strategy that aims to match the intended PT ($\text{PT}_q$) of the input query with the known PTs of the candidate brand entities. The PT information for a brand entity is pre-computed based on the catalog data. 

Since we do not have an authoritative source for brand-PT relationships, we approximate this mapping indirectly from products stamped with the brand entities. For every unique brand entity, we sample and collect a list of impressed products (those shown as part of search results for any query) over a four-month period, where their corresponding PTs are aggregated. Hence for each brand entity, we have a list of associated PTs. We note that this is not an exhaustive list of PTs and we are unable to collect this data for every brand entity. We use the Q2PT model to predict the product-type for the query. If the dictionary lexical matcher has mapped the brand surface form to multiple brand entities, we only select those brand entities that have PT matching the Q2PT prediction. We found that this filtering was able to narrow down the prediction to a single brand entity for over 50\% of multi-brand-entity prediction cases. We do not make any brand-entity prediction for the queries that still map to multiple brand entities to ensure high precision. 

\subsection{NER $+$ Exact Lexical Matching}
\label{ner_exact_matching}

In our first benchmark, we implement the matcher as a static dictionary with the keys being brand names and the values being the brand entities. For an input query, MetaTS-NER detects the brand name in its string form. The output of MetaTS-NER is then fed to the matcher. The dictionary serves as an exact string matcher that performs a lookup based on this output. We construct this dictionary from the internal brand-name2entity dataset (detailed in \ref{sec:dataset}). 
In the dictionary dataset, each brand entity may have multiple textual representations (e.g., full names and their abbreviated forms). We collect all valid representations as potential keys in our dictionary. 
Since some brand name representations are store-specific, we append the store tag to the beginning of the dictionary key. For representations that are common across multiple stores, we duplicate the same brand name and add corresponding store tag as prefix. While each brand entity is unique, brand names on the other hand do not have this constraint. 


\begin{figure*}[htbp]
  \centering
  \begin{subfigure}[b]{0.44\textwidth}
    \centering
    \resizebox{\columnwidth}{!}{\begin{tikzpicture}[
    node distance=1.6cm,
    every node/.style={font=\large},
    box/.style={rectangle, rounded corners, minimum width=2.5cm, minimum height=1.2cm, text centered, draw=black, fill=#1},
    oval/.style={ellipse, minimum width=2cm, minimum height=1cm, text centered, draw=black, fill=#1},
    arrow/.style={thick,->,>=stealth},
    process/.style={box=blue!20},
    process1/.style={box=blue!5},
    input/.style={oval=blue!40},
    output/.style={box=orange!30},
    module/.style={box=green!10}
]

\node[input] (query) {Query};
\node[module, right=of query] (pecos) {PECOS};
\node[module, right=of pecos] (filtering) {Filtering};
\node[output, right=of filtering, align=center] (output) {Brand Entity\\Prediction};

\node[process1, below=0.8cm of pecos, minimum width=3.5cm, align=center] (train) {Query $\rightarrow$ \\ Brand Entity \\ Training Data};
\node[process1, below=0.8cm of filtering, minimum width=4cm, align=center] (info) {Brand Entity \\ and Product \\ Type Associations};

\draw[arrow] (query) -- (pecos);
\draw[arrow] (pecos) -- (filtering);
\draw[arrow] (filtering) -- (output);
\draw[arrow] (train) -- (pecos);
\draw[arrow] (info) -- (filtering);

\draw[dashed, thick, blue, rounded corners] 
    ($(pecos.north west)+(-1.0,0.8)$) rectangle ($(filtering.south east)+(1.3,-2.6)$);

\node at ($(pecos)!0.5!(filtering)+(0,1.0)$) {\textbf{Q2E PECOS (Query2Entity PECOS)}};

\end{tikzpicture}}
    \caption{PECOS-based End-to-End framework.}
    \label{pecos_end_to_end}
  \end{subfigure}
  \hfill
  \begin{subfigure}[b]{0.50\textwidth}
    \centering
    \resizebox{\columnwidth}{!}{\begin{tikzpicture}[
    node distance=1.2cm,
    every node/.style={font=\Large},
    box/.style={rectangle, rounded corners, minimum width=2.5cm, minimum height=1.2cm, text centered, draw=black, fill=#1},
    oval/.style={ellipse, minimum width=2cm, minimum height=1cm, text centered, draw=black, fill=#1},
    arrow/.style={thick,->,>=stealth},
    process/.style={box=blue!20},
    input/.style={oval=blue!40},
    output/.style={box=orange!30},
    module/.style={box=green!10}
]

\node[input] (query) {Query};
\node[module, right=of query, yshift=2cm] (q2pt) {Q2PT};
\node[module, right=of query] (ner) {NER};
\node[module, right=of ner] (matcher) {Matcher};
\node[module, right=of query, yshift=-2cm] (q2e) {Q2E-PECOS};
\node[process, right=of matcher, yshift=-0.0cm, minimum width=3cm, minimum height=5cm, align=center] (post) {Post-\\Processing};
\node[output, right=of post, align=center] (output) {Brand Entity\\Prediction};

\draw[arrow] (query) -- (q2pt.west);
\draw[arrow] (query) -- (ner);
\draw[arrow] (query) -- (q2e.west);
\draw[arrow] (q2pt) -- (q2pt -| post.west);
\draw[arrow] (ner) -- (matcher);
\draw[arrow] (matcher) -- (post);
\draw[arrow] (q2e) -- (q2e -| post.west);
\draw[arrow] (post) -- (output);

\end{tikzpicture}}
    \caption{Fusion of one-stage and two-stage solutions.}
    \label{fig:fusion-solution}
  \end{subfigure}
  \vspace{-0.2cm}
  \caption{(a) A \textbf{PECOS} based End-to-End framework for brand entity linking. The framework maps queries to brand entities using extreme multi-class classification, followed by product type-based filtering. (b) Fusion of one-stage and two-stage approaches, combining their predictions with priority given to lexical matching results.}
  \label{fig:combined}
  \vspace{-0.2cm}
\end{figure*}
\subsection{NER $+$ PECOS Semantic Matching}
\label{subsec-ner-pecos}

Simple string matching based on surface forms often fails to capture brand name variations. When different textual representations of the same brand (such as abbreviations and full names) are not all included in the dictionary, some valid mentions may be missed. To address this limitation, we formulate the matching step as an eXtreme Multi-class Classification (XMC) problem and implement the matcher $g$ using PECOS~\cite{yu2022pecos}.
Given a brand mention $m$ extracted by MetaTS-NER, the XMC objective learns a scoring function $f: \mathcal{X} \rightarrow \mathbb{R}^L$ that maps the mention to a relevance score vector over $L$ brand entities ($L \approx 60{,}000$). PECOS decomposes this into a hierarchical structure:
\begin{equation}
g_{\text{M2E}}(m) = (E_{\text{ID}}, S), \quad S_i = \langle \mathbf{w}_i, \phi(m) \rangle
\end{equation}
where $\phi(m)$ is the TF-IDF feature representation of the mention and $\mathbf{w}_i$ is the learned weight vector for entity $i$. The hierarchical label tree enables inference in $O(b \log L)$ time with beam size $b$, making it suitable for real-time query understanding. The entity disambiguation step then leverages both the relevance scores and product type context:
\begin{equation}
e = h(E_{\text{ID}}, S, q, \text{PT}_q)
\end{equation}

\textbf{Model training and inference:} Unlike the lexical matcher which is parameter-free, PECOS requires training data consisting of (\textit{brand name, brand entity}) pairs. The model consumes the MetaTS-NER brand name prediction as input and classifies it into one of the brand entities. A single multilingual M2E-PECOS model handles all languages. When PT-based filtering cannot disambiguate, the relevance score $S$ is used to select the highest-scored entity.

\section{Brand Entity Linking via Extreme Multi-Class Classification Framework}
The two-stage approaches have the following limitations: i) Since the two-stage approach relies on two independent steps, errors can propagate from one step to the following step, ii) The first-stage NER creates a bottleneck for recall performance, i.e., we are unable to retrieve a brand entity if the MetaTS-NER model fails to recognize a brand. 
Furthermore, string matching can only retrieve explicitly mentioned brands from queries. Brands can also be implied through product names. For instance, a product line name may indicate a specific manufacturer without explicitly stating it. The two-stage approach cannot retrieve such implicit brand references.
To tackle these limitations, we propose an end-to-end method for brand entity linking that directly predicts brand entities from input queries.

For the end-to-end model, 
we trained a PECOS model that directly takes a query as input and predicts relevant brand entities. This simplifies model training, deployment, and inference significantly compared to the two-stage approach.
Mathematically, the end-to-end approach can be formulated as follows:
Given a query $q$, the Q2E-PECOS model directly predicts a set of candidate brand entities with their relevance scores, followed by a refinement step:
\begin{align}
    (E_{\text{ID}}, S) &= g_{Q2E}(q), \nonumber \\
    e &= h(E_{\text{ID}}, \text{PT}_q, S, q),
\end{align}
where $S$ contains the corresponding relevance scores for predicted entities $E_{\text{ID}}$. The filtering step utilizes the PECOS relevance scores along with PT-matching-based disambiguation. We refer to this model as Q2E-PECOS. The label space for Q2E-PECOS is collected from the brand-name2entity dataset (see \ref{sec:dataset}), i.e., 61,697 brand entities. The brand entity vocabulary is predefined and derived from the product catalog.

Note that many search queries do not contain a brand intent. The input to our brand entity linking model is all search queries, out of which branded queries are only a subset. To address this, we add a \textit{NIL} class to the brand entity space to allow our model to handle non-branded queries. In other words, we expect the Q2E-PECOS to predict \textit{NIL} label for a non-branded query.

\textbf{Model Training and Inference}: The Q2E-PECOS model directly maps queries to brand entities, which differs from the two-stage framework in Section~\ref{sec_two_stage_framework}. 
The training data of Q2E-PECOS is multilingual (detailed in \ref{sec:dataset}). The training data contains both valid brand entities and a dummy entity (\textit{NIL}).
In the inference stage, when multiple classes are predicted for an input query, \textit{NIL} (if predicted) and valid brand entities are ranked according to their relevance scores. A simple way is to choose the highest-scored class on the basis of PT match for the input query, which is what we implement in this approach.

\textbf{Complexity}: The computational complexity of this approach is determined by the PECOS framework. For a given input query, the time complexity is $O(b \times \log L)$, where $b$ is the beam size and $L$ is the number of brand entities. This is significantly more efficient than traditional approaches that have $O(L)$ complexity. The space complexity is also reduced from $O(d \times L)$ to $O(d \times \log L)$ where $d$ is the dimension of the feature vector.

\section{Fusion of the Q2E-PECOS and Exact Lexical Match Prediction}\label{fusion}

The proposed Q2E-PECOS model as a one-stage solution is simpler and capable of covering wider variety of brand names than exact string matching based method. However, while Q2E-PECOS allows us to retrieve various brands, it shows lower precision compared to the exact string matching based solution (as shall be discussed for results in Table~\ref{eval-table}). To receive the benefits of both methods, we can run the two methods in parallel independently from each other and fuse their predictions as a post-processing step to make the final prediction for a query. When both predictions are available for a query, we select the MetaTS-NER + exact lexical match prediction over Q2E-PECOS to opt for higher precision. The diagram of the fusion solution is shown in Figure~\ref{fig:fusion-solution}.

Given a query $q$, brand entity predictions by the (NER based) lexical matching approach and the Q2E-PECOS approach are executed in parallel. A priority-based fusion outputs the NER lexical prediction when present, deferring to Q2E-PECOS for queries where NER produces no valid brand entity. This approach allows us to leverage the high precision of the (NER-based) lexical matching method while benefiting from the higher coverage of the Q2E-PECOS method. The computational complexity of the fusion approach is the sum of the complexities of both methods, but since they can be run in parallel, the actual runtime is determined by the slower of the two methods.

\section{Experiments}
\subsection{Datasets}
\label{sec:dataset}
We use three training data sources of increasing supervision: Brand2entity (B2E), Strongly-labeled (SL), and Weakly-labeled (WL), summarized in Table~\ref{tab:dataset-stats}. Each is described in detail below.
\begin{table}[h]
  \caption{Dataset Statistics}
  \label{tab:dataset-stats}
  \vspace{-0.3cm}
  \centering
  \resizebox{\columnwidth}{!}{
  \begin{tabular}{cccccc}
    \toprule
    \textbf{\# Brand Entities} & \multicolumn{3}{c}{\textbf{Training}} & \textbf{Test} \\
    & Brand2entity & SL & WL & \\
    \midrule
    61,697 & 616,974 & 806,972 & 1,308,816 & 28,439 \\
    \bottomrule
  \end{tabular}
  }
\vspace{-0.3cm}
\end{table}

\subsubsection{Training Dataset}
The training dataset consists of three folds and is explained as the following.

\textit{Brand2entity}: Brand2entity is internal dictionary-like dataset with each entry being \textit{brand name, brand entity} pair. The dataset has two use cases: i) It serves as part of the training data where a brand name is considered as a pseudo query. ii) it is also used for building the strongly-labeled dataset. As for why we use this as part of training data, we consider it as a form of data augmentation. Similar augmentation has been explored in the extreme zero-shot text classification~\citep{xiong2022extreme} and large-scale retrieval of the search engine~\citep{tay2022transformer}. 



\textit{Strongly-labeled}: The strongly-labeled dataset is a dataset derived from search queries annotated by human judges internally. The data contains both branded as well as non-branded queries. 
The internal annotations consist of top query attributes like \textit{brand}, \textit{color}, \textit{size} etc. The data set is used for NER model training.
For the purpose of our brand entity linking, we only utilize the \textit{brand} labels and mark all other attributes as $O$ (Other). 
The dataset covers 13 languages: English, Spanish, German, French, Italian, Japanese, Dutch, Portuguese, Polish, Turkish, Swedish, Arabic and Czech. We collect 806972 training pairs from this dataset.
The dataset contains the brand name from the human annotated label. To build mapping from a search query to brand entity, we add an additional step of mapping the brand name to its corresponding brand entity through exact string matching by leveraging brand-name2entity dataset. This gives us the strongly-labeled dataset.


\textit{Weakly-labeled}: Although the brand-name2entity data source is rich in its brand distribution, the brand names alone do not match the query distribution. Since we ultimately wish to identify brand entities for search queries, we sampled and aggregated the historical search traffic of e-commerce platform to collect a weakly-labeled set of branded queries. We sampled query-product pairs that have a strong association through past search engagement aggregated over a time period. From the sampled data, we also collect the brand names for each of these products. We then check if the brand name is present in the associated query string as a substring. If it is present, we label the query with that brand name. Note that we rely on the strong query-product association to make this labeling and we do not verify these labels through human annotation. We build 1308816 such examples as part of training data.

\subsubsection{Test Dataset}

\begin{table}
  \caption{Percentage Distribution of Queries by Language}
  \label{tab:test_language_distribution}
  \vspace{-0.3cm}
  \resizebox{\columnwidth}{!}{
  \begin{tabular}{lccccccccccc}
    \toprule
    \textbf{Language} & EN & ES & DE & FR & JP & IT & AR & TR & NL & PT & SV \\
    \midrule
    \textbf{\%} & 39.6 & 12.0 & 11.7 & 9.4 & 6.5 & 5.5 & 5.0 & 4.6 & 2.6 & 2.2 & 1.1 \\
    \bottomrule
  \end{tabular}
  }
  \vspace{-0.3cm}
\end{table}

We use a held-out subset of the strongly-labeled dataset as our gold test dataset. As mentioned previously, the dataset contains queries with their annotated brand names and the associated entities. 
We add the corresponding brand entities through exact string matching using the brand name. The mapping process can sometimes lead to more than one brand entity being selected. Since we cannot be sure of the right one, we discard such examples for the purpose of precision and recall measurement. 
There are 28439 queries in total in the test dataset, out of which 24054 queries (84.5\%) are labeled with single brand entity whereas 4385 queries are multi-brand-entity labeled queries.
This test dataset represents query distribution across 11 languages, with English accounting for the largest share at nearly $40\%$ as shown in Table~\ref{tab:test_language_distribution}. The remaining languages are more evenly distributed, with Spanish, German, French, and Japanese making up most of the mid-range, while others like Swedish and Portuguese have relatively small shares. Overall, it provides a balanced multilingual benchmark for evaluating experimental performance across both high- and low-resource languages.

\subsection{Evaluation Metrics}
Following the typical Recall and Precision definition, we define our evaluation metrics as follows:
\begin{align*}
  \text{Recall} &= \frac{|\mathcal{C}|}{|\mathcal{L}_{\text{single}}|}, 
  \text{Precision} &= \frac{|\mathcal{C}|}{|\mathcal{P}_{\text{single}}|}, 
  \text{Coverage} &= \frac{|\mathcal{P}_{\text{single}}|}{|\mathcal{T}|},
\end{align*}
where $\mathcal{T}$ is the set of all test queries, $\mathcal{L}_{\text{single}} \subseteq \mathcal{T}$ is the subset of test queries with single-brand-entity labels, $\mathcal{P}_{\text{single}} \subseteq \mathcal{T}$ is the subset of test queries for which the model predicted a single-brand-entity,
and $\mathcal{C} = \{ q \in \mathcal{L}_{\text{single}} \cap \mathcal{P}_{\text{single}} \mid \text{pred}(q) = \text{label}(q) \}$ is the set of correctly predicted single-brand-entity queries.


The difference between the recall and coverage metrics is the following: i) recall only counts the correctly predicted single-brand-entity labeled query number whereas coverage includes also wrong single-brand-entity predicted cases as well. ii) As for denominators, recall counts the queries with single-brand-entity labels whereas
coverage also counts the non-single-brand-entity labeled queries. We include the coverage metric to account for measuring the valid predictions including both correct and incorrect single-label predictions. Apart from the above metrics, we also include the standard F1-score metric for the benchmarks. 
\subsection{Benchmarks}
We benchmarked multiple approaches for the brand entity linking task, covering both 
two-stage and end-to-end frameworks:

\begin{itemize}
    \item {NER + Exact Lexical Match:} A two-stage pipeline where MetaTS-NER 
    detects brand mentions, followed by exact dictionary-based matching to brand entities.  

    \item {NER + M2E-PECOS:} A two-stage approach where brand mentions from 
    MetaTS-NER are mapped to brand entities using PECOS, formulated as an extreme 
    multi-class classification problem.  

    \item {Bi-encoder:} A baseline entity linking model~\cite{li2020efficient} that encodes queries 
    and entities separately into embedding spaces, with similarity-based matching. We used Qwen3 Embedding 0.6B model~\cite{zhang2025qwen3embed} that is latest trained with larger capacity and better generalization ability to have a strong baseline. The B2E dataset is used for brand entity representation.

    \item {Q2E-PECOS:} An end-to-end PECOS-based framework that directly 
    predicts brand entities from queries without an intermediate NER stage.  

    \item {Fusion models:} Hybrid approaches that combine predictions from 
    the lexical matcher and a candidate model, leveraging the high precision of lexical matching 
    with the broader coverage of end-to-end prediction. It is indicated by the \checkmark notation in the Fusion column of Table~\ref{eval-table}
\end{itemize}

For both M2E-PECOS and Q2E-PECOS, the TF-IDF features are computed at the word level using unigrams (1 million) and bigrams (3 million) features.
Each method was trained under different dataset configurations: \textit{B2E} as Dictionary-based brand2entity pairs,  \textit{SL} as Strongly-labeled human-annotated queries, and \textit{WL} as Weakly-labeled queries derived from query--product interactions. The benchmark results under these models and training data settings are reported in Table~\ref{eval-table}.

\subsection{Numerical Results}

\begin{table*}
	\caption{Evaluation of the benchmark and proposed methods. All numbers are in percentage (\%). We use \best{bold} and \comparable{italic} fonts to highlight the best and 2nd best metric for each column, respectively. The \checkmark notation of Fusion column indicates the candidate model of a corresponding row is benchmarked with fusion strategy with baseline NER followed with exact lexical match.}
  	\label{eval-table}
    \vspace{-0.3cm}
   	\centering
	\resizebox{\textwidth}{!}{%
\begin{tabular}{ccccc cccc cccc}
   \toprule
   \textbf{Method} & \multicolumn{3}{c}{\textbf{Training Data}} & \textbf{Fusion} & \multicolumn{4}{c}{\textbf{Group-1\_AVG}} & \multicolumn{4}{c}{\textbf{Group-2\_AVG}} \\
	{}& B2E& SL & WL & {}& Coverage & Recall & Precision & F1 & Coverage & Recall & Precision & F1 \\
	\hline
  	{\begin{tabular}[x]{@{}c@{}} NER $+$ Exact \\ Lexical Match\end{tabular}}&
   \checkmark & \checkmark & {} & {} & 58.28 & 64.66 & \best{97.22} & 77.67 & 70.16 & 86.21 & \best{99.15} & 92.23 \\
  	\hline
	\multirow{3}{*}{\begin{tabular}[x]{@{}c@{}} NER $+$ \\ M2E-PECOS\end{tabular}} 
        & \checkmark & {} & {} & {} & 71.15 & 63.88 & 88.58 & 74.23 & \comparable{88.21} & 81.19 & 90.85 & 85.75 \\
        & \checkmark & \checkmark & {} & {} & 60.86 & 59.58 & 95.23 & 73.30 & 75.90 & 76.53 & 95.18 & 84.84 \\
        & \checkmark & \checkmark & {} & \checkmark & 67.51 & 67.64 & 96.05 & 79.38 & 83.31 & 88.57 & 98.73 & 93.37 \\
    \hline
    \multirow{2}{*}{Bi-encoder} & {\checkmark} & {} & {} & {} & 38.44 & 16.34 & 43.04 & {16.43} & 53.56 & 20.52 & 38.77 & {43.04} \\
    {} & {\checkmark} & {} & {} & \checkmark & {74.88} & 70.26 & 86.98 & {77.75} & {86.37} & 90.71 & 94.98 & {92.84} \\
    \hline
    
    \multirow{5}{*}{Q2E-PECOS} 
        & \checkmark & {} & {} & {} & \best{93.13} & 60.96 & 65.56 & 63.18 & \best{95.94} & 73.70 & 76.74 & 75.19 \\
    	& \checkmark & \checkmark & {} & {} & 49.68 & 50.56 & 94.89 & 65.97 & 63.97 & 72.40 & 97.76 & 83.19 \\
    	& \checkmark & \checkmark & {} & \checkmark & 70.98 & \comparable{75.26} & \comparable{96.13} & \comparable{84.42} & 80.77 & \best{94.71} & \comparable{98.92} & \best{96.77} \\
    	& \checkmark & \checkmark & \checkmark & {} & 62.50 & 62.14 & 93.46 & 74.65 & 73.35 & 78.76 & 96.97 & 86.92 \\
    	& \checkmark & \checkmark & \checkmark & \checkmark & \comparable{75.31} & \best{77.35} & 94.93 & \best{85.24} & {85.09} & \comparable{94.64} & 98.55 & \comparable{96.56} \\
	\bottomrule
\end{tabular}
	}
\end{table*}

We present our experimental results in Table \ref{eval-table}. We present multiple experiment runs depending on the training data and fusion configurations used. Here, the fusion parameter refers to post-processing fusion strategy mentioned in Section~\ref{fusion}.

We first measure precision, recall and coverage metrics for each online retail store. 
The evaluation is performed over two store clusters, i.e. Group-1 and Group-2. 
Group-1 includes 9 marketplaces covering English, German, French, Italian, Spanish, and Japanese with higher traffic volume. Group-2 includes 9 marketplaces covering Portuguese, Turkish, Arabic, Dutch, and Swedish in addition to English and Spanish variants, representing lower-resource settings with less training data available.
The two groups differ greatly in size, evaluation data volume and search query pattern distributions. Hence, we average the metrics for both groups separately.

We first look at the two-stage methods, i.e. MetaTS-NER $+$ Exact Lexical Match and M2E-PECOS. With the fusion technique, it can be seen that semantic matching via PECOS does help to improve the coverage over the exact lexical matching technique. But we see a drop in both recall and precision when moving from exact lexical match to semantic match. While the semantic matching technique can retrieve more brand entities, it does so with a lower precision. When a fusion of the lexical and semantic matching techniques is applied, we see improvement in both coverage and recall, whereas the precision is slightly affected.


We also evaluate a Bi-encoder baseline and a Bi-encoder + fusion variant (Table~\ref{eval-table}). The standalone Bi-encoder shows substantially lower recall and coverage than both the two-stage methods and the PECOS-based models, which we attribute to the difficulty of learning reliable semantic similarity from very short, noisy queries and the long-tailed brand distribution. Applying the same fusion strategy (i.e., giving priority to high-precision NER + exact lexical matches and falling back to the Bi-encoder predictions) improves coverage and recall relative to the standalone Bi-encoder — demonstrating that lexical matching can compensate for many missed semantic matches. However, even with fusion the Bi-encoder’s precision and overall F1 remain lower than those achieved by PECOS-based models with fusion, indicating that PECOS better balances coverage and precision for the extreme, sparse label space encountered in e-commerce brand linking.


Next, we compare the Q2E-PECOS model to the baseline model MetaTS-NER $+$ Exact Lexical Match. We see that the training data choice has a significant impact on the performance of the Q2E-PECOS model. This is expected since the end-to-end model requires a lot more data to learn the task. When only using the brand-name2entity and the strongly-labeled datasets, Q2E-PECOS performs worse than the MetaTS-NER + Exact Lexical Matching model across all metrics. Adding the fusion strategy improves the model performance significantly. This shows that with limited data, both models have complementary effects. When we add in the weakly-labeled catalog data, we see the Q2E-PECOS model improve over our baselines even without the fusion mechanism. Finally, we get the best performance across all metrics when combining all training data sources and also including the fusion mechanism.

\begin{table}[h]
  \centering
  \caption{False Alarm Rate (\%) on Non-branded Queries}
  \label{false-alarm}
  \vspace{-0.3cm}
  \begin{tabular}{lc}
    \toprule
    \textbf{Model} & \textbf{False Alarm Rate (\%)} \\
    \midrule
    NER $+$ Exact Lexical Match & 1.177 \\
    NER $+$ M2E-PECOS & 3.267 \\
    Q2E-PECOS w/o WL Data & 4.037 \\
    Q2E-PECOS w/ WL Data & 6.550 \\
    \bottomrule
  \end{tabular}
  \vspace{-0.3cm}
\end{table}



\textbf{False alarm evaluation:} All metrics shown in Table \ref{eval-table} were measured on the gold test set containing only branded queries. In the real-world e-commerce services, we wish to deploy such a model for all queries and expect the model to be able to identify branded queries and predict the corresponding brand entity. To study the impact of the model in such a setting, we collected $\sim$85K non-branded queries and observed the predictions by the different models. We define the false alarm rate in this evaluation as the percentage of non-branded queries for which the model incorrectly predicts a brand entity. Table \ref{false-alarm} reports the rates for the four main models. As we can see, the exact lexical match based method receives the lowest false alarm. The M2E-PECOS has a slightly larger false alarm and Q2E-PECOS model further increases the rate by $\sim$3\%. Although we see an increased false alarm rate with our best Q2E-PECOS model, we believe the increase is still acceptable given the overall performance gain of this model along with its more simplified implementation.

\textbf{Error analysis:} To understand the remaining failure modes, we manually categorized 5{,}876 mismatch cases from the two-stage baseline across five major marketplaces. The dominant error type is NER detection failure (73.1\% of errors), split between cases where the brand name is lexically present in the query but undetected (33.6\%) and cases involving abbreviations or implicit references (39.5\%, e.g., product model codes that require world knowledge). The remaining 26.9\% are entity linking errors: parent/sub-brand confusion (11.4\%), spelling and formatting variants (3.4\%), and disambiguation failures in multi-brand queries or product-line names (12.1\%). This breakdown motivates the fusion approach, as the end-to-end Q2E-PECOS model bypasses NER entirely and directly recovers many of the 73.1\% detection-stage failures.

\section{Online Results from A/B Test}
\begin{table}[h]
\centering
\caption{Online A/B Test Comparison: Control (NER + Lexical Match) vs. Treatment Q2E-PECOS with Fusion Solution. All reported gains are statistically significant.}
\label{tab:abtest-results}
\vspace{-0.3cm}
\resizebox{\columnwidth}{!}{
\begin{tabular}{lcc}
\hline
\textbf{Metric} & \textbf{Group-1 Stores} & \textbf{Group-2 Stores} \\
\hline
Brand Entity Recall & +11.00\% & +5.44\% \\
Customer Engagement & \multicolumn{2}{c}{+0.02\%} \\
Immediate Contribution Profit & \multicolumn{2}{c}{+0.03\%} \\
\hline
\end{tabular}
}
\end{table}

Identifying brands and linking them to the correct entities is a critical component of query understanding in e-commerce search. Accurate brand recognition ensures that relevant products are retrieved and ranked appropriately, directly influencing customer experience. Errors in brand entity linking can lead to irrelevant results, reduced engagement, and a negative shopping experience.

\paragraph{Online A/B Test.}
To evaluate the effectiveness of our proposed methods, we conducted a two-week online A/B test under real-world conditions across 11 languages. The control system used the NER model followed by exact lexical matching, while the treatment system employed the fusion solution with Q2E-PECOS (Section~\ref{fusion}).

The results in Table~\ref{tab:abtest-results} demonstrate that the fusion solution improves brand entity recall by +11.0\% in Group-1 cluster stores and $+5.44\%$ in Group-2 cluster stores. Beyond recall, the online experiment shows statistically significant improvements in downstream business metrics: customer engagement increased by +0.02\%, contribution profit improved by +0.03\%, and sponsored brand revenue grew by +0.06\%. These gains confirm that more robust brand entity linking directly translates to improved product retrieval, ranking relevance, and business outcomes across multiple application surfaces including organic search, sponsored products, and brand-specific merchandising.

\paragraph{Computational Cost.}
The PECOS-based models achieve sub-5ms latency per query due to their hierarchical sparse structure, making them suitable for real-time search. The NER component (DistilBERT~\citep{sanh2019distilbert}) adds $\sim$10ms. The fusion approach runs both pipelines in parallel, with total latency bounded by the slower NER path ($\sim$10ms). In contrast, the bi-encoder baseline requires $\sim$52ms per query due to dense embedding generation, exceeding typical production latency budgets. This motivates our choice of PECOS over dense retrieval approaches. We note that autoregressive entity linking systems (e.g., GENRE~\citep{cao2021autoregressive}, mGENRE~\citep{de2022multilingual}) and cross-encoder approaches (e.g., BLINK~\citep{wu2020scalable}) require 50--100ms+ per query and assume Wikipedia-style entity descriptions unavailable in our proprietary brand catalog, making them incompatible with our real-time constraints and domain.

\section{Conclusion}
In this work, we proposed a two-stage method and an end-to-end classification approach for brand entity linking for e-commerce search queries. We conduct extensive experiments using various modeling and data augmentation techniques and show improved performance compared to the more commonly used two-stage approaches to this problem. We also explore the feasibility of deploying such a model on real-world search query traffic. Given the result, we show that there is good scope for improvement in exploring end-to-end techniques for attribute extraction tasks. We also built a fusion solution including both the two-stage and one-stage models into one framework. The solution has been validated by an online A/B test.

\section*{GenAI Usage Disclosure}
Generative AI tools were used solely for manuscript polishing (grammar and clarity improvements). No AI tools were used in the research methodology, experimental design, data collection, or analysis.

\bibliographystyle{ACM-Reference-Format}
\balance
\bibliography{mybib}

@inproceedings{kolitsas2018end,
	author = {Kolitsas, Nikolaos and Ganea, Octavian-Eugen and Hofmann, Thomas},
	booktitle = {Proceedings of the 22nd Conference on Computational Natural Language Learning (CoNLL)},
	pages = {519--529},
	doi = {10.18653/v1/K18-1050},
	title = {End-to-end neural entity linking},
	year = {2018}}

@inproceedings{sorokin2018mixing,
	author = {Sorokin, Daniil and Gurevych, Iryna},
	booktitle = {Proceedings of the Seventh Joint Conference on Lexical and Computational Semantics (*SEM)},
	pages = {65--75},
	doi = {10.18653/v1/S18-2007},
	title = {Mixing context granularities for improved entity linking on question answering data across entity categories},
	year = {2018}}

@inproceedings{peters2019knowledge,
	author = {Peters, Matthew E and Neumann, Mark and Logan, Robert and Schwartz, Roy and Joshi, Vidur and Singh, Sameer and Smith, Noah A},
	booktitle = {Proceedings of the 2019 Conference on Empirical Methods in Natural Language Processing and the 9th International Joint Conference on Natural Language Processing (EMNLP-IJCNLP)},
	pages = {43--54},
	doi = {10.18653/v1/D19-1005},
	title = {Knowledge enhanced contextual word representations},
	year = {2019}}

@article{yu2022pecos,
	author = {Yu, Hsiang-Fu and Zhong, Kai and Zhang, Jiong and Chang, Wei-Cheng and Dhillon, Inderjit S},
	journal = {Journal of Machine Learning Research},
	number = {98},
	pages = {1--32},
	title = {{PECOS}: Prediction for enormous and correlated output spaces},
	volume = {23},
	year = {2022}}

@inproceedings{chang2021extreme,
	author = {Chang, Wei-Cheng and Jiang, Daniel and Yu, Hsiang-Fu and Teo, Choon-Hui and Zhang, Jiong and Zhong, Kai and Kolluri, Kedarnath and Hu, Qie and Shandilya, Nikhil and Ievgrafov, Vyacheslav and Singh, Japinder and Dhillon, Inderjit S},
	booktitle = {Proceedings of the 27th ACM SIGKDD International Conference on Knowledge Discovery \& Data Mining},
	title = {Extreme Multi-label Learning for Semantic Matching in Product Search},
	year = {2021}}

@inproceedings{etter2021accelerating,
	author = {Etter, Philip A and Zhong, Kai and Yu, Hsiang-Fu and Ying, Lexing and Dhillon, Inderjit},
	booktitle = {Proceedings of the Web Conference},
	title = {Accelerating Inference for Sparse Extreme Multi-Label Ranking Trees},
	year = {2022}}

@inproceedings{xiong2022extreme,
	author = {Xiong, Yuanhao and Chang, Wei-Cheng and Hsieh, Cho-Jui and Yu, Hsiang-Fu and Dhillon, Inderjit},
	booktitle = {NAACL},
	title = {Extreme Zero-Shot Learning for Extreme Text Classification},
	year = {2022}}

@inproceedings{tay2022transformer,
	author = {Tay, Yi and Tran, Vinh Q and Dehghani, Mostafa and Ni, Jianmo and Bahri, Dara and Mehta, Harsh and Qin, Zhen and Hui, Kai and Zhao, Zhe and Gupta, Jai and others},
	booktitle = {Advances in Neural Information Processing Systems (NeurIPS)},
	volume = {35},
	title = {Transformer memory as a differentiable search index},
	year = {2022}}

@inproceedings{le2019distant,
	address = {Florence, Italy},
	author = {Le, Phong and Titov, Ivan},
	booktitle = {Proceedings of the 57th Annual Meeting of the Association for Computational Linguistics},
	doi = {10.18653/v1/P19-1400},
	month = jul,
	pages = {4081--4090},
	publisher = {Association for Computational Linguistics},
	title = {Distant Learning for Entity Linking with Automatic Noise Detection},
	url = {https://aclanthology.org/P19-1400},
	year = {2019}}

@inproceedings{zwicklbauer2016robust,
	address = {New York, NY, USA},
	author = {Zwicklbauer, Stefan and Seifert, Christin and Granitzer, Michael},
	booktitle = {Proceedings of the 39th International ACM SIGIR Conference on Research and Development in Information Retrieval},
	doi = {10.1145/2911451.2911535},
	isbn = {9781450340694},
	location = {Pisa, Italy},
	numpages = {10},
	pages = {425--434},
	publisher = {Association for Computing Machinery},
	series = {SIGIR '16},
	title = {Robust and Collective Entity Disambiguation through Semantic Embeddings},
	url = {https://doi.org/10.1145/2911451.2911535},
	year = {2016}}

@inproceedings{morene2017combine,
	address = {Cham},
	author = {Moreno, Jose G. and Besan{\c{c}}on, Romaric and Beaumont, Romain and D'hondt, Eva and Ligozat, Anne-Laure and Rosset, Sophie and Tannier, Xavier and Grau, Brigitte},
	booktitle = {The Semantic Web},
	editor = {Blomqvist, Eva and Maynard, Diana and Gangemi, Aldo and Hoekstra, Rinke and Hitzler, Pascal and Hartig, Olaf},
	isbn = {978-3-319-58068-5},
	pages = {337--352},
	publisher = {Springer International Publishing},
	title = {Combining Word and Entity Embeddings for Entity Linking},
	year = {2017}}

@inproceedings{zhang2019joint,
	address = {New York, NY, USA},
	author = {Fang, Zheng and Cao, Yanan and Li, Qian and Zhang, Dongjie and Zhang, Zhenyu and Liu, Yanbing},
	booktitle = {The World Wide Web Conference},
	doi = {10.1145/3308558.3313517},
	isbn = {9781450366748},
	location = {San Francisco, CA, USA},
	numpages = {10},
	pages = {438--447},
	publisher = {Association for Computing Machinery},
	series = {WWW '19},
	title = {Joint Entity Linking with Deep Reinforcement Learning},
	url = {https://doi.org/10.1145/3308558.3313517},
	year = {2019}}

@inproceedings{spitkovsky12a,
	address = {Istanbul, Turkey},
	author = {Spitkovsky, Valentin I. and Chang, Angel X.},
	booktitle = {Proceedings of the Eighth International Conference on Language Resources and Evaluation ({LREC}'12)},
	month = may,
	pages = {3168--3175},
	publisher = {European Language Resources Association (ELRA)},
	title = {A Cross-Lingual Dictionary for {E}nglish {W}ikipedia Concepts},
	url = {http://www.lrec-conf.org/proceedings/lrec2012/pdf/266_Paper.pdf},
	year = {2012}}

@inproceedings{logeswaran-etal-2019-zero,
	author = {Logeswaran, Lajanugen and Chang, Ming-Wei and Lee, Kenton and Toutanova, Kristina and Devlin, Jacob and Lee, Honglak},
	booktitle = {Proceedings of the 57th Annual Meeting of the Association for Computational Linguistics},
	title = {Zero-Shot Entity Linking by Reading Entity Descriptions},
	year = {2019}}

@article{huang2015leverageing,
	author = {Hongzhao Huang and Larry P. Heck and Heng Ji},
	bibsource = {dblp computer science bibliography, https://dblp.org},
	eprint = {1504.07678},
	eprinttype = {arXiv},
	journal = {CoRR},
	title = {Leveraging Deep Neural Networks and Knowledge Graphs for Entity Disambiguation},
	url = {http://arxiv.org/abs/1504.07678},
	volume = {abs/1504.07678},
	year = {2015}}

@inproceedings{cao-etal-2017-bridge,
	address = {Vancouver, Canada},
	author = {Cao, Yixin and Huang, Lifu and Ji, Heng and Chen, Xu and Li, Juanzi},
	booktitle = {Proceedings of the 55th Annual Meeting of the Association for Computational Linguistics (Volume 1: Long Papers)},
	doi = {10.18653/v1/P17-1149},
	month = jul,
	pages = {1623--1633},
	publisher = {Association for Computational Linguistics},
	title = {Bridge Text and Knowledge by Learning Multi-Prototype Entity Mention Embedding},
	url = {https://aclanthology.org/P17-1149},
	year = {2017}}

@inproceedings{Li2021Meta,
	author = {Zheng Li and Danqing Zhang and Tianyu Cao and Ying Wei and Yiwei Song and Bing Yin},
	booktitle = {EMNLP 2021},
	title = {MetaTS: Meta teacher-student network for multilingual sequence labeling with minimal supervision},
	url = {https://www.amazon.science/publications/metats-meta-teacher-student-network-for-multilingual-sequence-labeling-with-minimal-supervision},
	year = {2021}}

@article{sevgili2020neural,
	author = {Sevgili, {\"{O}}zge and Shelmanov, Artem and Arkhipov, Mikhail and Panchenko, Alexander and Biemann, Chris},
	journal = {Semantic Web},
	volume = {13},
	number = {3},
	pages = {527--570},
	title = {Neural Entity Linking: {A} Survey of Models based on Deep Learning},
	year = {2022}}

@article{reinanda2020kg,
	author = {Ridho Reinanda and Edgar Meij and Maarten de Rijke},
	doi = {10.1561/1500000063},
	issn = {1554-0669},
	journal = {Foundations and Trends{\textregistered} in Information Retrieval},
	number = {4},
	pages = {289-444},
	title = {Knowledge Graphs: An Information Retrieval Perspective},
	url = {http://dx.doi.org/10.1561/1500000063},
	volume = {14},
	year = {2020}}

@inproceedings{cao2021autoregressive,
  author    = {Nicola {De Cao} and Gautier Izacard and Sebastian Riedel and Fabio Petroni},
  title     = {Autoregressive Entity Retrieval},
  booktitle = {9th International Conference on Learning Representations ({ICLR})},
  publisher = {OpenReview.net},
  year      = {2021},
  url       = {https://openreview.net/forum?id=5k8F6UU39V}}

@inproceedings{li2020efficient,
 title={Efficient One-Pass End-to-End Entity Linking for Questions},
 author={Li, Belinda Z. and Min, Sewon and Iyer, Srinivasan and Mehdad, Yashar and Yih, Wen-tau},
 booktitle={EMNLP},
 year={2020}
}

@misc{zhang2025qwen3embed,
      title={Qwen3 Embedding: Advancing Text Embedding and Reranking Through Foundation Models}, 
      author={Yanzhao Zhang and Mingxin Li and Dingkun Long and Xin Zhang and Huan Lin and Baosong Yang and Pengjun Xie and An Yang and Dayiheng Liu and Junyang Lin and Fei Huang and Jingren Zhou},
      year={2025},
      eprint={2506.05176},
      archivePrefix={arXiv},
      primaryClass={cs.CL},
      url={https://arxiv.org/abs/2506.05176}, 
}

@article{Sukumaretal2024,
  author = {Sushmi Thushara Sukumar and Chung-Horng Lung and M. Zaman and Ritesh Panday},
  title = {Knowledge Graph Generation and Application for Unstructured Data Using Data Processing Pipeline},
  year = {2024},
  journal = {IEEE Access},
}

@article{Tanetal2017,
  author = {Chuanqi Tan and Furu Wei and Pengjie Ren and Weifeng Lv and M. Zhou},
  title = {Entity Linking for Queries by Searching Wikipedia Sentences},
  year = {2017},
  journal = {Conference on Empirical Methods in Natural Language Processing},
}

@article{Manchandaetal2020,
  author = {Saurav Manchanda and Mohit Sharma and G. Karypis},
  title = {Distant-Supervised Slot-Filling for E-Commerce Queries},
  year = {2020},
  journal = {2021 IEEE International Conference on Big Data (Big Data)},
}

@article{Zhangetal2021,
  author = {Danqing Zhang and Zheng Li and Tianyu Cao and Chen Luo and Tony Wu and Hanqing Lu and Yiwei Song and Bing Yin and Tuo Zhao and Qiang Yang},
  title = {QUEACO: Borrowing Treasures from Weakly-labeled Behavior Data for Query Attribute Value Extraction},
  year = {2021},
  journal = {International Conference on Information and Knowledge Management},
}

@article{Ricatteetal2023,
  author = {Thomas Ricatte and Donato Crisostomi},
  title = {AVEN-GR: Attribute Value Extraction and Normalization using product GRaphs},
  year = {2023},
  journal = {Annual Meeting of the Association for Computational Linguistics},
}

@article{Luoetal2024,
  author = {Chen Luo and Xianfeng Tang and Hanqing Lu and Yaochen Xie and Hui Liu and Zhenwei Dai and Limeng Cui and Ashutosh Joshi and Sreyashi Nag and Yang Li and Zhen Li and R. Goutam and Jiliang Tang and Haiyang Zhang and Qi He},
  title = {Exploring Query Understanding for Amazon Product Search},
  year = {2024},
  journal = {BigData Congress [Services Society]},
}

@article{Kozarevaetal2016,
  author = {Zornitsa Kozareva and Qi Li and Ke Zhai and Weiwei Guo},
  title = {Recognizing Salient Entities in Shopping Queries},
  year = {2016},
  journal = {Annual Meeting of the Association for Computational Linguistics},
}

@article{Manshadietal2009,
  author = {Mehdi Manshadi and Xiao Li},
  title = {Semantic Tagging of Web Search Queries},
  year = {2009},
  journal = {Annual Meeting of the Association for Computational Linguistics},
}

@article{Wenetal2019,
  author = {Musen Wen and D. Vasthimal and Alan Lu and Tiantian Wang and Aimin Guo},
  title = {Building Large-Scale Deep Learning System for Entity Recognition in E-Commerce Search},
  year = {2019},
  journal = {BDCAT},
}

@article{GuisadoGamezetal2021,
  author = {Joan Guisado-Gámez and David Tamayo-Domènech and J. Urmeneta and J. Larriba-Pey},
  title = {ENRICH: A Query Rewriting Service Powered by Wikipedia Graph Structure},
  year = {2021},
  journal = {Wiki@ICWSM},
}

@article{de2022multilingual,
  title={Multilingual Autoregressive Entity Linking},
  author={De Cao, Nicola and Wu, Ledell and Popat, Kashyap and Artetxe, Mikel and Goyal, Naman and Plekhanov, Mikhail and Zettlemoyer, Luke and Cancedda, Nicola and Riedel, Sebastian and Petroni, Fabio},
  journal={Transactions of the Association for Computational Linguistics},
  volume={10},
  pages={274--290},
  year={2022},
  publisher={MIT Press},
  doi={10.1162/tacl_a_00460}
}

@inproceedings{wu2020scalable,
  title={Scalable Zero-shot Entity Linking with Dense Entity Retrieval},
  author={Wu, Ledell and Petroni, Fabio and Josifoski, Martin and Riedel, Sebastian and Zettlemoyer, Luke},
  booktitle={Proceedings of the 2020 Conference on Empirical Methods in Natural Language Processing (EMNLP)},
  pages={6397--6407},
  year={2020}
}

@inproceedings{sanh2019distilbert,
  title={DistilBERT, a distilled version of BERT: smaller, faster, cheaper and lighter},
  author={Sanh, Victor and Debut, Lysandre and Chaumond, Julien and Wolf, Thomas},
  booktitle={5th Workshop on Energy Efficient Machine Learning and Cognitive Computing at NeurIPS},
  year={2019}
}

@inproceedings{kroll2024rexel,
  title={REXEL: An End-to-end Model for Document-Level Relation Extraction and Entity Linking},
  author={Kroll, Theodor Axel and Eberts, Markus and Tran, Nhat-Khang and Stein, Anja and Klinger, Roman},
  booktitle={Proceedings of the 2024 Conference of the North American Chapter of the Association for Computational Linguistics (NAACL): Industry Track},
  pages={123--136},
  year={2024}
}

@inproceedings{wang2025aelc,
  title={AELC: Adaptive Entity Linking with LLM-Driven Contextualization},
  author={Wang, Yiming and Gao, Xuan and Huang, Shuxin and Li, Zeyu and Cao, Jianbo and Huo, Jidong},
  booktitle={Findings of the Association for Computational Linguistics: EMNLP 2025},
  year={2025}
}

@inproceedings{nag2023qau,
  title={Query Attribute Understanding},
  author={Nag, Sreyashi and Liu, Dong and Gupta, Grishma and Sunkara, Monica and Avudaiappan, Neela},
  booktitle={Proceedings of the 61st Annual Meeting of the Association for Computational Linguistics (ACL): Industry Track},
  pages={144--155},
  year={2023}
}

@inproceedings{chen2022queryattr,
  title={Query Attribute Recommendation at Amazon Search},
  author={Chen, Chen and Liu, Hanqing and Xiao, Bingqing and Lu, Hanqing and He, Qi},
  booktitle={Proceedings of the 45th International ACM SIGIR Conference on Research and Development in Information Retrieval: Industry Track},
  year={2022}
}

\end{document}